\documentclass[preprint,tightenlines,eqsecnum,aps,prd,nofootinbib]{revtex4}

\usepackage{subfigure}%
\usepackage{graphicx}%
\usepackage{enumerate}%
\usepackage{amsmath}%
\usepackage{amsfonts}

\def\be{\begin{equation}}
\def\bea{\begin{eqnarray}}
\def\ee{\end{equation}}
\def\eea{\end{eqnarray}}
\def\aux { [ \! ]   }
\def\pip { |   }

\def\H { {\cal H}   }
\def\openone{\leavevmode\hbox{\small1\kern-3.3pt\normalsize1}}%
\def\Rds{R}
\def\roughly#1{\raise.3ex\hbox{$#1$\kern-.75em\lower1ex\hbox{$\sim$}}}

\tighten

\begin{document}
\title{A global picture of quantum de Sitter space}

\author{Steven B. Giddings and Donald Marolf}

\affiliation{Department of Physics, University of California,   \\
Santa Barbara, CA 93106, USA \\ \texttt{giddings@physics.ucsb.edu,
marolf@physics.ucsb.edu}}

\begin{abstract}
Perturbative gravity about a de Sitter background motivates a global
picture of quantum dynamics in `eternal de Sitter space,'  the
theory of states which are asymptotically de Sitter to both future
and past.   Eternal de Sitter physics is described by a finite
dimensional Hilbert space in which {\it each} state is precisely
invariant under the full de Sitter group.  This resolves a
previously-noted tension between de Sitter symmetry and finite
entropy.  Observables,  implications for Boltzmann brains, and
Poincar\'e recurrences are briefly discussed.
\end{abstract}

\date{May 2007}
 \maketitle

\section{Introduction}
\label{intro}

\medskip

Modern cosmology provides ample motivation to understand de
Sitter-like cosmologies in detail.   The success of early universe
inflation and current observations suggest that our universe emerged
from an early high-scale de Sitter-like stage and is now at the
onset of a low-scale stage. A feature of particular interest is the
finite Bekenstein-Hawking entropy $S_{dS}$ of the de Sitter (dS)
horizon, which raises many questions about de Sitter microstates. It
is often hoped that an understanding of this deep issue will have
practical implications for the thorny problem of extracting
predictions from a theory with eternal inflation,  e.g. in the
context of the string theory landscape.

Our work probes in such directions by exploring the perhaps simpler physics of `eternal de Sitter space,'  by which we mean the theory of states which are asymptotically de Sitter to both future and past with the {\it same} cosmological constant $\Lambda$.  Our approach will be to investigate {\it perturbative} gravity about a de Sitter background in the regime where such a theory is self-consistent, and to use the results to motivate a deeper picture of eternal de Sitter quantum gravity.

Eternal de Sitter quantum gravity has been previously investigated
by many authors.  While there is no complete consensus, a common
feature of many analyses has been to emphasize the region, centered
on some freely falling worldline, in which de Sitter space may be
described as a static spacetime. These analyses also emphasize the
`static Hamiltonian' which generates time translations in this
`static patch.'  This emphasis has led various authors to suggest
that dS physics is a close analogue to a static finite box at
roughly the dS temperature $T_{dS}$, but with a finite number of
states \cite{Banks:2000fe,Fischunpub,Dyson:2002nt} $\exp(S_{dS})$.

However, it is difficult to find a rationale for such a static
hamiltonian to be physical; certainly in a perturbative approach it
appears unphysical. The reason for this is as follows.  At zeroth
order, the theory is just matter plus free gravitons on a de Sitter
background.  In this context, consider some `boost' symmetry $B$ of
dS, which generates a corresponding symmetry of the zeroth order
theory.  In $D$ spacetime dimensions, $B$ fixes an $S^{D-2}$ on the
equator of $S^{D-1}$, and the Killing horizons of $B$ are the
cosmological horizons of two observers which we take to lie at the
north and south poles of $S^{D-1}$ as shown in figure 1. The boost $B$
generates a symmetry of the order-zero theory, and this generator
may be written as $B= H_S - H_N$, where $H_N, H_S$ are the so-called
static Hamiltonians of the associated north and south static
patches.

\begin{center}
\includegraphics[width=3.5cm]{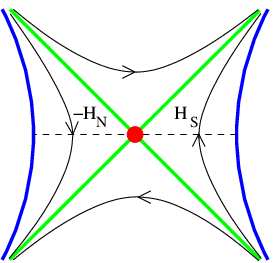} 
\end{center}
FIG.  1: De Sitter space with boost Killing field $B =  H_S - H_N$ and horizons.  The dashed line is the `neck,'  the smallest $S^{D-1}$  in this frame.  The dot is the $S^{D-2}$ left fixed by $B$.  The worldline of the south (north) pole observer is the right (left) edge of the diagram. 
\medskip

The picture then changes at first non-trivial order in gravitational interactions.  Here, so called linearization stability constraints \cite{VMLS,Arms,VM,FM,AMM,AH1,LU} arise, which state that only zeroth order states satisfying $B=0$ lead to consistent first order perturbations.  The statement is akin to the fact that the total electric charge on a compact space must vanish.  The point is that, at this (first)  order, the {\it zeroth}-order boost generator $B$ coincides with one of the gravitational Hamiltonian constraints and must vanish identically.     While the action of $B$
translates the static patch in time, we see that $B$ generates a
gauge symmetry and not a physically interesting notion of time
evolution.  In a perturbative quantum treatment, one requires that $B$ annihilate
physical states, so that the two static Hamiltonians $H_S$ and $H_N$ are perfectly correlated.

What do these facts imply for the action of $H_S$?   We may consider
two contexts.  First, as has become common, let us confine attention
to the south patch and trace over degrees of freedom in the north.
But due to the above correlations with $H_N$, the resulting density
matrix $\rho_S$ is diagonal in a basis of $H_S$-eigenstates. Thus,
$[\rho_S,H_S]=0$ and $H_S$ does not generate an useful notion of
time evolution.

One might also consider the action of $H_S$ on a region which overlaps the north patch.  In doing so, we note that $H_S$ alone does not
generate a de Sitter symmetry. Instead, it generates the analogue of boosting the
right Rindler wedge of Minkowski space while keeping the rest of Minkowski
space fixed. Indeed, acting with $H_S$ alone changes the short
distance structure of the vacuum near the equatorial $S^{D-2}$. As a
result, any action of  $H_S$ on such a region falls outside the scope of low energy physics.  We conclude that the static Hamiltonian does
not provide a useful basis for an analysis of de Sitter physics\footnote{An alternate interpretation is that  a physical static Hamiltonian does exist, but that it is a genuinely new ingredient, visible only in the full quantum theory.  This appears to be the viewpoint of \cite{BanksStable}, whose notion of a ``static Hamiltonian'' has eigenvalues bounded by the de Sitter temperature $T_{dS}$.  This operator vanishes in the classical limit, and at this level is consistent with the arguments above.  However, there remains some tension with the fact that such an operator does not appear at the semiclassical level, where $T_{dS} \neq 0$ .}.  

In order to capture the effects noted above, we continue to address
de Sitter physics below using perturbative gravity about the full de
Sitter background.  Our main focus (section \ref{GA}) is on the
perturbative state space, but we also briefly address perturbative
observables in section \ref{obs}. We take a global perspective, but
confine ourselves to contexts where gravitational back-reaction is
small.  The above-mentioned linearization stability constraints play
a key role in our analysis.  Section \ref{picture} then describes a
picture of eternal de Sitter in full quantum gravity which is
motivated by the results of sections \ref{GA} and \ref{obs}.    Some
remaining issues are discussed in section \ref{disc}.  We plan to
supply further details in future work.

\section{Eternal de Sitter states}
\label{GA}

As noted in the introduction, the fact that Cauchy slices of de
Sitter space are compact (spheres) leads to an interesting
constraint on physical states, even when gravity is treated
perturbatively.  We explore the implications of such constraints
below, and argue that eternal de Sitter physics is described by a
finite-dimensional space in which each state is precisely de Sitter
invariant.

\subsection{de Sitter invariant states}

Our context here is perturbative gravity.  To be specific, consider the expansion of a gravitating theory in inverse powers of the Planck mass $m_p$, in which we take the graviton to have canonical normalization.  At order zero, we have free gravitons together with a matter quantum field theory on a fixed de Sitter background.  The matter theory may contain non-gravitational interactions of finite strength.  For simplicity, we restrict attention to matter fields having a dS-invariant vacuum.

This order-zero theory has a large state space ${\cal H}_0$, but
only certain states have consistent perturbative coupling to the
gravitational field.  At first order in $1/m_p$, one finds
`linearization stability constraints'
\cite{VMLS,Arms,VM,FM,AMM,AH1,LU} forcing each de Sitter charge
$Q[\xi] = \int_\Sigma T_{ab} n^a \xi^b$ to vanish. Here $\xi$ is a
de Sitter Killing field, $\Sigma$ is a Cauchy surface, and $Q[\xi]$
generates the corresponding isometry.   Thus, only dS-invariant
quantum states couple consistently to perturbative gravity; we refer to such
states as `physical' states.  Similarly, only dS-invariant operators
are observables. This is a direct analogue of a familiar feature of
Maxwell gauge fields, where the presence of a compact Cauchy surface
requires the total electric charge of any matter state to vanish.
Thus, in the Maxwell context, allowed matter states are U(1)
invariant, and only U(1) invariant operators will preserve this
space of states.

While the order-zero Hilbert space (${\cal H}_0$) may have many
states with zero electric charge, typically only the vacuum state
will be fully de Sitter invariant \cite{AHII}.  As suggested in
\cite{AHII}, one may nevertheless proceed by considering formally
dS-invariant sums of states in ${\cal H}_0$, so long as one
introduces a `renormalized' inner product on the space of such sums.
In particular,  for each state $\aux \psi \rangle$ in ${\cal H}_0$,
consider the expression \be \label{ga} \pip \Psi \rangle :=   \eta
\aux \psi \rangle :=  \int dg U(g)  \aux \psi \rangle, \ee where
integral is over the de Sitter group and $dg$ is the standard Haar
measure.  The result clearly satisfies the physical state
condition\footnote{The formal expression (\ref{ga}) for $\pip \Psi
\rangle$ may be given meaning as a `distributional' state analogous
to the delta-function position or momentum eigenstates of quantum
mechanics. In particular, for appropriate $\aux \psi \rangle$,
expression (\ref{ga}) yields a well-defined linear function on an
appropriate subspace of ${\cal H}_0$. We consider only states $\aux
\psi \rangle$ for which (\ref{ga}) converges in this sense.  The
dual action of $Q[\xi]$ on $\pip \Psi \rangle$ then annihilates
$\pip \Psi \rangle$.  See e.g. \cite{ALMMT}.}  $Q[\xi] \pip \Psi
\rangle =0$. We will use the symbol $\eta$ below to denote this
`group averaging' operation.

For physical states of the form (\ref{ga}), consider the new inner product:
\be
\label{gaip}
\langle \Psi_1 \pip \Psi_2 \rangle:=
 \langle \psi_1 \aux   \int dg U(g)  \aux \psi_2 \rangle.
 \ee
For certain exactly solveable free theories, \cite{AHII,AHprivate}
showed that (\ref{gaip}) is finite when $\aux \psi_1 \rangle, \aux
\psi_2 \rangle$ are Fock basis states containing a sufficient number
of particles\footnote{The norm (\ref{gaip}) is clearly {\it not}
finite for  a dS-invariant vacuum state $\aux 0 \rangle$, which must
be treated separately.  Such a separate treatment is justified by
the fact that, for a well-defined observable ${\cal O}$, the state ${\cal O}\aux0\rangle$ is dS-invariant and normalizeable in $\H_0$; i.e., it is proportional to $\aux 0 \rangle$.  Thus $\langle \psi \aux {\cal O} \aux 0\rangle$ vanishes whenever $\aux 0\rangle$ and $\aux  \psi \rangle$ are
orthogonal.  (To avoid later confusion, we note that the action of (\ref{sint}) on $\aux 0 \rangle$ does not yield a normalizeable state in $\H_0$; the action of (\ref{sint}) is  well-defined only on the orthogonal complement of $\aux 0 \rangle$.) See
\cite{ALMMT,single,SO21,JL} for comments on this issue. 
We are concerned here with generic states, and so will not
treat the vacuum in detail. }  (typically 2 or 4). When completed
with respect to the inner product (\ref{gaip}), the space of
group-averaged states  forms the `physical' Hilbert space ${\cal
H}_{phys}$.  This construction was also introduced independently in
the general theory of constrained systems \cite{KL} and in the
context of quantum cosmology \cite{QORD}.  The reader may consult
the by now extensive literature (e.g.
\cite{ALMMT,single,WhereAreWe,MG,OS}) for more details.  One may
also motivate this approach via the functional integral.  (See e.g.
\cite{GMH,DMPI} for related discussions.)

For (\ref{gaip}) to be a valid inner product, it must be positive
semi-definite.   Ref. \cite{AHII} showed that this is the case for a
1+1 free scalar toy model and for 3+1 linearized gravitons. While
examples (in other contexts) are known for which group averaging is
not positive definite \cite{JL}, these cases are rather singular
even at the classical level.  Furthermore, when group averaging
converges, one can show \cite{MG}  that  it  gives the unique
physical inner product consistent with the *-algebra of observables
on ${\cal H}_0$; i.e., if it fails to be positive definite, then
{\it no} renormalized inner product will be positive.   Below, we
shall simply assume that (\ref{gaip}) is positive when it converges,
though an argument in appendix \ref{count} below shows that
(\ref{gaip}) is indeed positive on a `large' class of states.

\subsection{States with weak back-reaction}

Perturbation theory can be valid only when gravitational interactions are weak.  We should therefore confine attention to a appropriate subspace ${\cal H}^w \subset {\cal H}_0$ of states which meet this criterion.    Note that while such an ${\cal H}^w$ will be de Sitter invariant as a linear space, the individual states in ${\cal H}^w$ will not be dS-invariant.  Group averaging will be required to transform them into physical states.

A variety of gravitational effects must be controlled for a state to
lie in ${\cal H}^w$.  For example, two energetic particles passing
close to each other can result in a large gravitational scattering
angle or in the formation of a black hole.  Black hole formation may
also result from the collective interaction of a large number of
particles localized on a scale small compared to the de Sitter
length scale $\Rds$.  However, these sorts of large gravitational
effects also happen in the asymptotically flat context. On the other
hand, a  new gravitational effect for dS is that matter spread over
an entire Cauchy surface  can cause the overall acceleration of the
universe to slow, and can even cause the spacetime to undergo a `big
crunch.'  It is on this latter sort of effect that we wish to focus.

For the moment, let us consider
a state $\aux \psi \rangle \in {\cal H}_0$ of the matter field which is rotationally invariant under some $SO(D)$ subgroup of the $SO(D,1)$ de Sitter group.  This rotational symmetry picks out  a preferred foliation of dS by spheres of symmetry.  At early times the spheres are large and contract rapidly.  The contraction gradually slows so that the spheres reach a minimum `neck' of radius $r=\Rds$ and then begin to reexpand.   For such states,  the greatest danger of large gravitational back-reaction generically occurs at this neck, and the size of the gravitational back-reaction is controlled by the total flux of energy passing through the neck.  

Let us therefore fix a foliation ${\cal F}$ of dS by round spheres
and consider the energy flux operator
 \be
  \label{fluxdef} F = \int_{neck} \sqrt{{}^{\scriptscriptstyle D-1}g} \ T_{ab} n^a n^b ,
 \ee
where the integral is over the neck determined by the given
foliation, ${}^{\scriptscriptstyle D-1}g$ is the determinant of the
induced metric, $n^a$ is the unit future-pointing normal to the
neck, and $T_{ab}$ is a renormalized stress tensor such that
$\langle T_{ab} \rangle =0$ in the dS-invariant vacuum.   $F$ is a
well-defined operator with a discrete spectrum, free of accumulation
points\footnote{This claim can be verified directly for free fields,
and for conformally invariant matter by conformally mapping $F$ (up
to a c-number anomaly term) to the Hamiltonian on the Einstein
static universe.  One expects it to hold in general.}.  Let us also
choose a cut-off $f$, such that a uniformly distributed energy
density with $F < f$ has small gravitational back-reaction.
Dimensional analysis provides a rough estimate of the value of such
$f$ as the appropriate power of the de Sitter radius,
 \be
  \label{fb}
 f\sim \Rds^{D-3}/G\quad ,
 \ee
where $G \sim 1/m_p^{D-2}$.     Since the spectrum of $F$ has no
accumulation points, for a given foliation $\cal F$ the bound $F<f$
will be satisfied only by a finite number of quantum states.
Because generic states with $F<f$ are nearly uniform, and because we
are most interested in the collective gravitational effects on the
spacetime as a whole, we will neglect the fact that non-uniform
states with $F<f$ may still have large back-reaction.
The states ${\cal H}^{{\cal F},f}$ having $F < f$ with respect to the given foliation ${\cal F}$
(modulo those that are locally strongly coupled), comprise a
subspace of ${\cal H}^w$.

On the other hand, there are many states in ${\cal H}^w$ having
large $F$, greater than $f$. In particular, the definition of $F$
required a choice of foliation by spheres.  Since the possible
foliations are related by de Sitter transformations, which in
general include de Sitter boosts, a state which has small $F$ with
respect to one foliation may have large $F$ with respect to another.
Since the de Sitter group is a gauge group in perturbative gravity,
such states must also be considered to have small back-reaction.
Physically, one notes that in a flat background the Aichelburg-Sexl
metric is believed to give a largely accurate description of a
highly-boosted particle, and one expects the same of the de Sitter
version of this metric \cite{Hotta:1992qy,Esposito:2006vv}.

We therefore take a state to lie in ${\cal H}^w$ if it has $F < f$
with respect to some foliation ${\cal F}$.  In fact, ${\cal H}^w$
will contain  the space ${\cal H}^{f} = \oplus_{\cal F} \  {\cal
H}^{{\cal F}, f}$ generated through arbitrary (normalizeable)
superpositions of states in the individual ${\cal H}^{{\cal F}, f}$
for various foliations ${\cal F}$.   Since one expects ${\cal
H}^{f}$ to include all states with weak back-reaction, we will use
${\cal H}^{f}$ as a proxy for ${\cal H}^w$ below.

Due to the freedom to choose any foliation ${\cal F}$, the space
${\cal H}^{f}$ has infinite dimension.  But the foliations differ
only by dS transformations $g$, and group averaging takes states
$\aux \psi_1 \rangle$ and $U(g) \aux \psi_1 \rangle$ to the same
physical state.  The dimension of the corresponding {\it physical}
space ${\cal H}^{f}_{phys} = \eta ({\cal H}^{f})$ is thus no larger
than that of a single  ${\cal H}^{{\cal F}, f}$ for a fixed
foliation.  As stated above, this dimension is finite.  In fact,
using the techniques of \cite{stiff} and assuming a typical
relativistic equation of state $p=\frac{\rho}{D-1}$,  the logarithm
of the dimension of ${\cal H}^{{\cal F}, f}$ may be estimated
roughly as $(\Rds/\ell_p)^{(D-1)(D-2)/D}$, where $\ell_p$ is the
Planck length.    In appendix A, it is shown that group averaging
does not significantly change this result; i.e., the logarithm of
the dimension of $\H^f_{phys}$ is also of this order.

To understand the full Hilbert space of physical states requires a
complete theory of dS quantum gravity.  While this is beyond the
scope of our work, we note that the dimension of ${\cal
H}^{f}_{phys}$ is consistent with the conjecture
\cite{Banks:2000fe,Fischunpub} that the full space of asymptotically
(past and future) de Sitter states is of finite dimension, with
entropy given by the Bekenstein-Hawking value $S_{dS} =
(\Rds/\ell_p)^{D-2}$.  It is not a surprise that the number of
weakly-coupled states is smaller than the de Sitter entropy.  The
argument is a standard one: At least when the number of species is
small, the entropy of a given region of space is maximized when a
black hole is present, which by definition is not a weakly-coupled
phenomenon.  To obtain  the space of all asymptotically de Sitter
states,  we should include black holes (and possibly other
non-perturbative gravitational states) up to the maximum size $\sim
\Rds$.  The corresponding space of states ${\cal H}^{dS}$ thus
requires for its definition the full non-perturbative theory of
gravity, and is expected to have dimension roughly $e^{S_{dS}}$.

To summarize, we have identified a space ${\cal H}^f_{phys}$ for
which perturbative gravity is weakly coupled.  The construction of
${\cal H}^f_{phys}$ proceeds by 1) beginning with the original Fock
space ${\cal H}_0$ describing the limit in which gravity is
decoupled 2) restricting to the subspace ${\cal H}^{f}$ satisfying
$F<f$ with respect to {\it some} foliation $\cal F$ and 3) group
averaging to produce the space of states ${\cal H}^f_{phys}$.  The
logarithm of the dimension of ${\cal H}^f_{phys}$ is the size
expected from local field theory, and is less than $S_{dS}$.

\subsection{Future (and past) asymptotically de Sitter states}

While the space $\H^{dS}$  may be the complete space of states
describing a geometry that is asymptotically de Sitter to the past
and future, in describing observables and measurement it appears
useful to consider a much larger space.  Specifically, if we are
given a particular $\pip \Psi\rangle \in {\cal H}^{dS}$, physical
predictions involve computing amplitudes of the form $\langle
\Psi'|\Psi\rangle$ for the state $|\Psi\rangle$ to `look' like the
particular `descriptor state\footnote{An alternate perspective is
to take the descriptor states to be $\aux \psi' \rangle \in \H_0$.
Recalling that physical states  $\pip \Psi \rangle$ are best thought
of as functionals on (a subspace of) $\H_0$, the desired amplitude
$\langle \Psi \pip \Psi' \rangle$ is just the action of $\pip \Psi
\rangle$ on $\aux \psi' \rangle$. This action admits a perturbative
calculation if $\aux \psi' \rangle$ is weakly coupled on some Cauchy
surface.}' $\langle \Psi' |$.    For example,  $\langle\Psi'|$ might
describe the situation where you find yourself to be reading this
particular sentence, as opposed to something else, at a particular
instant in the evolution of the universe.  Alternatively,
$\langle\Psi'|$ might describe fluctuations of a scalar field of
some given magnitude, on an expanding spherical hypersurface with
radius $a$ satisfying $a \gg \Rds$.

The reason this leads to a space larger than ${\cal H}^{dS}$ is that
 a projection onto such a state does not in general commute with the requirement that
the universe be both past and future asymptotically de Sitter. Let
us consider the 2nd example above, and begin to evolve the
descriptor state backwards in time from the given hypersurface.  The
fluctuations then blueshift by a large factor, and generically lead
to a large gravitational back-reaction.  Thus it is natural to think
of  $\langle \Psi' |$ as lying outside ${\cal H}^{dS}$.  We refer to
the space of states described by data on a large expanding Cauchy
surface as the `future asymptotically de Sitter' space ${\cal
H}^{AsdS+}$, and one may analogously define ${\cal H}^{AsdS-}$.
Certain states in these enlarged spaces are also expected to play a
role in describing physical cosmologies that are only asymptotically
de Sitter in one direction.

Of course, many states in ${\cal H}^{AsdS\pm}$ define the same
functional on ${\cal H}^{dS}$.  Since ${\cal H}^{dS}$ is a Hilbert
space, the independent functionals on $\H^{dS}$ may be described as
the subset  ${\cal H}^{dS} \subset {\cal H}^{AsdS\pm}$.  We will
return to a discussion of ${\cal H}^{AsdS\pm}$ in section
\ref{picture}, but for use in section \ref{obs} we note that one may
wish to restrict ${\cal H}^{AsdS\pm}$ to various other subspaces as
well.   For example, recall that the cutoff $F < f$ leads to the
finite-dimensional subspace ${\cal H}^f_{phys} \subset  {\cal
H}^{AsdS+}$.  Likewise, to extend beyond this subspace, we could
consider applying a flux bound on the flux $F$ of (\ref{fluxdef})
computed through an $S^{D-1}$ at some radius $a$ much larger than
the neck radius, or with a different value of $f$.  For given $a$
and bound $F<f$ this defines a new space, ${\cal H}^{a,f}\subset
{\cal H}^{AsdS+}$, which is again of finite dimension.   Note that
there are inclusions ${\cal H}^{a,f}\subset {\cal H}^{a',f'}$ for
$a<a'$, $f<f'$.

\section{Observables, cut-offs, and locality}
\label{obs}

Though it is not our main focus here, it is appropriate to comment
on how familiar (i.e., local) physics is to be recovered from the de
Sitter-invariant physical states.   Recall that this de Sitter
invariance is a remnant of the diffeomorphism-invariance of the full
gravitational theory.  As a result, one expects  the interesting
dynamics to be {\it relational}
\cite{EinsDov,DeWitt:1962cg,DeWitt:1967yk,Page:1983uc,Banks:1984cw,Hartle:1986gn,Rovelli:1990jm,Rovelli:1990ph,Rovelli:1990pi,Rovelli:2001bz},
and in particular to describe, e.g., the {\it relative} positions of
various features of the quantum state.  In the quantum dS context,
this feature was pointed out in the original work \cite{VM}, where
the dS-invariance of states was first described.

Specifically, we expect to recover dynamics by using relational
observables (see e.g.
\cite{Rovelli:1990jm,Rovelli:1990ph,Rovelli:1990pi,Rovelli:2001bz,Tsamis:1989yu}\cite{QORD}\cite{MarT,Ambjorn:1996mk,Ambjorn:1996wc}\cite{GMH};
see also
\cite{Gambini:2004pe,Gambini:2006yj,Dittrich:2005kc,Thiemann:2004wk,Pons:2005rz}).
We refer the reader to \cite{GMH} for a discussion of how, at least
in appropriate states, `single integral observables' of the form
\begin{equation}
\label{sint}
{\cal O} = \int_{M} d^D x \sqrt{-g} A(x),
\end{equation}
allow one to approximately recover local physics.  An explicit example of this
procedure was provided in \cite{Gary:2006mw} for a two-dimensional
toy model, and general classes of one-dimensional models were
analyzed in \cite{QORD,MarT}.  Here $M$ is the full spacetime
manifold and $A(x)$ is a local scalar. The resulting ${\cal O}$ is
diffeomorphism-invariant and thus, in our context, de Sitter
invariant.   When $A(x) = C(x)D(x)$, one may think of such
observables as searching over all of $M$ to find an appropriate
detector $D(x)$ at which the observable reports the value of some
other field $C(x)$.  If several detectors are found, the values of
$C(x)$ at each detector are summed.

One would like to call ${\cal O}$ an ``observable.'' However,
 \cite{GMH} pointed out a problem: perturbative quantum gravity about a dS background is {\it not} a context in which ${\cal O}$ is a well-defined operator on either ${\cal H}_0$ or ${\cal H}_{phys}$.  Let us consider the case of  ${\cal H}_0$, which is somewhat simpler.  While it is straightforward to choose $A(x)$ for which typical matrix elements of ${\cal O}$ are finite, higher correlators involving ${\cal O}$ diverge.  Consider for example
\begin{equation}
\langle \psi_1 \aux  {\cal O} {\cal O} \aux \psi_2 \rangle = \int d^Dx \sqrt{-g} \int d^Dy \sqrt{-g}
\langle \psi_1 \aux  A(x) A(y)  \aux \psi_2 \rangle.
\end{equation}
Over most of the integration region, the correlator $\langle \psi_1
\aux  A(x) A(y)  \aux \psi_2 \rangle$ can be well approximated by
its vacuum value.  But the vacuum correlator is dS-invariant.  Thus,
the integrand does not change under the simultaneous action of a
dS-translation on both $x$ and $y$.   The integral over this `center
of mass' direction diverges.  In the `detector' interpretation
described above, the divergence is due to the small probability that
vacuum fluctuations give rise to virtual detectors in each small
region of space, over a spacetime of infinite volume.

Recalling that quantum fluctuations in the dS-invariant vacuum are thermal fluctuations in any static patch, such fluctuations may be picturesquely thought of as due to the
`Boltzmann brain' phenomenon
\cite{Boltzmann,Rees:1998jp,Albrecht:2004ke,Page:2006dt}, where a
detector or observer can thermally fluctuate into existence and make an
observation.  Formally, such `virtual' observers  can confound an
attempt to extract observations corresponding to detectors/observers
not produced by this mechanism, and this may present an ultimate
limit on constructing certain observables.

However, the cutoff prescriptions of the preceding subsection show
how to  regulate this divergence in a natural way.  Since matrix
elements of ${\cal O}$ are finite, the correlators diverge only
because an infinite number of intermediate states contribute.  But
our physical spaces $\H^{a,f}_{phys}$ have only finitely many
states.  The problem is thus that ${\cal O}$ does not preserve a
given $\H^{a,f}_{phys}$.  This happens because when $A(x)$ acts far
to the future of any dS-neck, it greatly changes the eigenvalue of
the associated flux operator $F$.  To remove the above divergence,
let $P^{a,f}_{phys}$ be the projector from $\H^{AsdS+}_{phys}$ to
$\H^{a,f}_{phys}$.  Then  $\tilde {\cal O} = P^{a,f}_{phys} {\cal O}
P^{a,f}_{phys}$ has well-defined correlators on $\H^{a,f}_{phys}$.
The projection $P^{a,f}_{phys}$ effectively limits the fluctuations
to a finite integration volume and so, for sufficiently small $a$
and $f$, ensures that rare vacuum fluctuations (aka Boltzmann
brains) give a negligible contribution.

This ameliorates the paradoxical results of
\cite{Page:2006dt,BF}, in the spirit of \cite{Hartle:2007zv}. Having
removed the problematic divergences, the stage is now set for the
recovery of local physics (in an appropriate limit) along the lines
described in e.g. \cite{GMH}.  More details of this limit will be
investigated in future work.

One may ask how large the parameter $a$ may be taken or,
equivalently, up to what  time past the neck such local observables may be
recovered. To estimate this limit, note that the fluctuations into
`virtual' observers are suppressed by the observer's mass $m$. The
probability for such a fluctuation is proportional to $
\exp(-m/T_{dS})$, where $T_{dS} \sim 1/\Rds$ is the de Sitter
temperature.  Any such observer should be contained in a causal
patch, and thus the largest such observer should have mass bounded
by $f \sim m_p^{D-2}R^{D-3}$. At time scales $t\roughly> \Rds
S_{dS}$, the exponentially growing volume of the global picture
produces a population explosion of such observers.  We thus find
that on these time scales, observations of even the largest
(and thus heaviest) conceivable observers can become confounded by this
miasma of Boltzmann brains. For small weakly coupled observers, the actual time
scale is strictly less, though it can still be quite large when the
cosmological constant is small.

\section{The Global Picture of Eternal de Sitter space}
\label{picture}

Thus far, we have studied in detail the states of perturbative
gravity about a de Sitter background which are weakly coupled over
the entire spacetime.   Restriction to weak coupling imposed a
cutoff on the energy flux through the de Sitter neck and resulted in
a finite-dimensional space of physical states ${\cal H}^f_{phys}$,
in which each state is precisely de Sitter invariant.  The logarithm
of the dimension of this space agress to leading order with the
entropy expected from local field theory in the absence of gravity.

We have also constructed observables on this space, from which local
dynamics are to be recovered.  The observables are relational, and
their construction raised ``Boltzmann brain'' issues which required a (relational) infra-red cutoff.
This cutoff is logically independent of the limitation of physical states to ${\cal H}^f_{phys}$, and we argued that it could be associated with larger spaces of  ``descriptor states'' such as  ${\cal H}^{a,f}_{phys}$

Which of these conclusions can be expected to hold in a more complete theory of eternal de Sitter quantum gravity?  We now address each point in turn.

 {\bf De Sitter invariant states:}  In the perturbative context, we emphasized that the de Sitter group acts as a {\it gauge} symmetry; it is part of the diffeomorphism gauge symmetry of low energy gravity and represents a redundancy in our description of the physics.  In a full theory of eternal de Sitter, such spacetime concepts may or may not play a fundamental role.  However, to the extent that any state may be said to be ``asymptotically de Sitter to the past and future,'' some notion of
 asymptotic de Sitter symmetry might be expected to be meaningful, and should again represent a redundancy of the description.  We therefore might expect physical states $\H^{dS}_{phys}$ of the full theory  to be in an appropriate sense  de Sitter invariant.

{\bf De Sitter invariant observables:}  By the same reasoning and to
the same extent, observables  of the full theory must also be de
Sitter invariant.  In this sense, meaningful dynamics will again be
fully relational, and local observables will approximately arise
from relational observables.  For more discussion, see \cite{GMH}.

{\bf The bound $F < f$:}  For states satisfying this bound (and
which are also locally weakly-coupled), perturbation theory is
weakly coupled everywhere in the spacetime. It is therefore
plausible that such states approximate states of the full theory,
though  we will return to this issue below.  On the other hand, as
has been pointed out by various authors (see e.g.
\cite{Banks:2000fe}), significant violations of this bound should
lead to gravitational collapse, where non-perturbative gravity is
clearly relevant.  It is an open question what the correct physics
is in this regime. A proposed  `nonlocality principle'
\cite{Giddings:2006sj,Giddings:2006be} suggests that the
non-perturbative gravitational dynamics of this regime has
essentially nonlocal large-scale behavior.  This is supported by
studies of high-energy scattering \cite{Giddings:2006vu} and by
investigating locality through relational observables
\cite{GMH,Gary:2006mw}, and could play an important role in
explaining unitarity of black hole evaporation
\cite{Giddings:2006sj,Giddings:2006be,Giddings:2007ie}. In the
present context the boundary $F\sim f$ in the field theory Fock
space would represent a correspondence point where local field
theory yields to such a new nonlocal theory needed for complete
description of the quantum dynamics.  In a flat
background, this conjectured correspondence point is parametrized by
the `locality bound' proposed for two-particle states in
\cite{Giddings:2006be,Giddings:2001pt,Giddings:2004ud} and in the N-particle case in
\cite{Giddings:2006vu}.  If this perspective is correct, our bound
$F<f$ is thus naturally termed a `de Sitter locality bound.'
Holographic information bounds \cite{dSAdv} have also been believed
to point to where local quantum field theory breaks down.
Assuming the expected connection between one bit of information and one quantum of energy
apparently
indicates that such bounds follow from locality bounds like our
bound $F<f$, but the latter bounds appear more general in that they
also apply to states with low entropy but high energy.

 {\bf Finite-dimensional state space:}  We have constructed a
space of states $\H^f_{phys}$ which we expect are perturbatively
close to actual interacting states corresponding to de Sitter space
and, at least for a small number of species, we have argued that the
number of such states is bounded by $e^{S_{dS}}$.  These states are
plausibly completed by nonperturbative states, involving black holes
and other strong-gravitational effects, to give a space ${\cal
H}^{dS}$ of quantum de Sitter states, whose dimension is $\sim
 e^{S_{dS}}$.  See \cite{Banks:2000fe,Fischunpub,dSAdv} for related discussion.

{\bf Large boosts and  long times:}  The group-averaging formalism
requires integration over the full non-compact de Sitter group. This
necessarily involves very large boosts, and the reader may ask if
this computation can be controlled in perturbative gravity even for
weakly coupled states in $\H^f_{phys}$.  Indeed,  we find it
plausible that subtle effects can be important in understanding the
detailed action of large boosts.  However, for given states
satisfying $F < f$, the perturbative theory suggests that the
contribution to the group averaging inner product falls off
exponentially quickly in the boost parameter.  In flat space, this
is just the statement that two states of the same mass are
orthogonal when their center-of-mass momenta differ by a large
boost.  This statement is expected to hold in the full theory, well
beyond the perturbative regime.  We will assume that it holds in the
full theory of eternal dS as well.

It is important to ask  in what regimes one might expect
significant deviations from the results of local field theory.
 One place it is reasonable to expect deviations to arise is
  at the point where the observables of local
field theory can no longer be recovered as limits of relational
observables in the full theory.   Our study of perturbative
observables suggested one such limitation, where consideration of
volumes larger than $\Rds^{D} e^{S_{dS}}$ led to typical observables
corresponding to even the largest observers being confounded by
vacuum fluctuations (`Boltzmann brains').  This suggests that local
field theory may break down for a complete description of global dS
for times longer than $t \sim \Rds S_{dS}$.

Other arguments  have also identified this timescale as signaling a
breakdown of local physics.   Specifically, in parallel to the
suggested
breakdown\cite{Page:1993wv}\cite{Giddings:2006sj,Giddings:2006be,Giddings:2007ie} of
local field theory in black hole contexts,
\cite{NAH} suggested that a global description breaks down at a time
scale $t\sim \Rds S_{dS}$.  Ref.~\cite{Giddings:2007ie} outlined an
argument for breakdown due to significant fluctuations on this time
scale, and \cite{NAH,Arkani-Hamed:2007ky} argued that this timescale
also appears in an attempt to regulate inflation by exiting a slow
rolling phase.  This same time scale also arose in \cite{DO}.

Together, these observations all point to the limit $t<\Rds S_{dS}$
on the time for which a local field theory description of the global
geometry makes sense.  At longer time-scales there may well be a
valid description of a smaller region of de Sitter, {\it e.g.} a
causal patch.  Indeed, confounding by Boltzmann brains grows more
probable with volume, so if one  restricts to a
causal patch, it appears that this effect becomes important only
on a time scale $t\sim \Rds e^{S_{dS}}$, which accords with
observations of \cite{dStrouble,Banks:2002wr} based on other
considerations.

In either context, recall from section \ref{obs} that observables
$\tilde {\cal O} = P^{a,f}_{phys} {\cal O} P^{a,f}_{phys}$ are
insensitive to such effects for moderate $a$, and plausibly for $a
\ll \Rds \exp(S_{dS})$.  Similarly, group averaging calculations for
states in $\H^f_{phys}$, and even in the larger spaces  of
descriptor states  $\H^{a,f}_{phys}$, are insensitive to
such effects for these values of $a$.  We therefore expect
computations of simple such correlators and matrix elements
performed in the local field theory approximation (with
gravitational perturbations) to reliably approximate results in
the full theory.  A similar conclusion should hold over similar
spacetime volumes of more general shapes.

\section{Discussion}
\label{disc}

A study of perturbative quantum gravity about a de Sitter background
has motivated the global picture of eternal de Sitter space
described above.  Key features are a finite number of physical
states, de Sitter invariance of {\it each} such state, relational
dynamics, and an outline of the recovery of local dynamics as a
limit for  times $t \ll \Rds S_{dS}$.  Recovery over longer
timescales may also be possible in limited regions of spacetimes,
such as within a causal patch.  The details of the local limit will
be explored in future work.

One notes that our construction provides a counter-example to the
claim of \cite{dStrouble} that the de Sitter symmetries are not
consistent with a finite entropy.  Which of their assumptions fail?
It turns out that it is their description of the states and
their dynamics.   In \cite{dStrouble} the contradiction arose from assuming that the dS generators act nontrivially on the finite number of physical states which were taken to be states of a single causal patch; the global picture was only used as a ``thermofield double'' formalism.
But the contradiction disappears if the de Sitter generators annihilate the correct physical states, as we have argued.

One may note that a na\"\i ve perturbative approach based on local
QFT still leads to a reduced density matrix $\rho_S$  (describing
e.g., the ``south'' causal patch) for which $- Tr [\rho \ln \rho]$
is formally infinite. But such a computation has nothing to do with
the dimension of our space of physical states.  In particular,
dynamics on our physical space is captured only by relational
observables with appropriate cutoffs. Relational observables are
also de Sitter invariant, and are not precisely restricted to the
static patch surrounding a particular geodesic. Now,  familiar local
physics on a fixed background can often be recovered  from
relational observables by taking an appropriate limit.   However, as
argued in e.g. \cite{GMH}, a precise  limit requires taking $\ell_p$
to zero, so that $S_{dS}$ diverges.   One expects that any attempt
to use relational observables to detect the entropy described in
\cite{dStrouble} while keeping $\ell_p$ fixed would be subject to a
cut-off which renders the entropy finite.\footnote{In the black
hole context, another physical motivation for a cut-off rendering
this entropy finite was given in \cite{Fr,RS,width}.}  In the global
picture, rather than finding a tension between finite entropy and
dS-invariance, we find that dS-invariance (and the associated
restriction to relational observables)  cuts off the entropy at a
finite value.

In short, it appears that some of the previously noted difficulties
arising in a causal-patch picture of de Sitter space are artifacts
 of that picture.  Another classic such issue is that of
Poincar\'e recurrences.  Since any time-translation invariant system
with a finite number of states $N \sim e^S$ experiences recurrences
over long times of order $e^S$, it was argued in
\cite{Dyson:2002nt,Dyson:2002pf} that similar issues should arise in
the static patch of de Sitter.  Now, as noted in
\cite{Banks:2002wr}, the operational status of such recurrences is
unclear as, within this finite dimensional state space, it is is not
possible to construct detectors, memory devices, etc. that can
meaningfully compare events separated on such long time scales.  The
detectors, memory devices, etc. will fall apart, stop working, or be
destroyed much earlier.  In addition, as we have described, on such
timescales observables describing local measurements are
confounded by fluctuations (section \ref{obs}). Nevertheless, there
is at least a sense in which, despite the finite dimension of our
state space, such recurrences do {\it not} arise at all in our
setup.  The point is that our dynamics is fully relational, and in
particular that it is natural to specify dynamics relative to
certain {\it global} features, such as presence of a `neck,' i.e., a
smallest $S^{D-1}$, or a larger $S^{D-1}$ with given radius $a$.
While global features experience quantum fluctuations, a thermal
description is not appropriate; global features do not recur per se.
Furthermore, dynamics specified relative to these global features
effectively takes place on a time-dependent background; e.g., space
is small at the neck but becomes arbitrarily large to both the
future and past. In such a strongly time-dependent setting, with
large asymptotic geometry, there is no reason for recurrences to
arise.

Of course, by studying de Sitter space in a
global description, we are working in a framework not apparently
obeying the strongest forms of the complementarity conjecture.  In
those forms, this conjecture has been taken to imply that only a
finite number of degrees of freedom describing a static patch make
sense, and that physics outside this region is not described by
independent degrees of freedom.   Our picture is
different.

Finally, we note that our considerations extend to portions of de
Sitter space that are either produced through tunneling transitions
or through prior FRW cosmologies, or which exit through termination
of inflation. Though the discussion of constraints in such contexts
is more subtle, dynamics will again be described by relational
observables.  Again, we expect to be able to effectively constrain
to consideration of observables in a limited region, {\it e.g.}
sufficiently near the region of transition into or out of de Sitter
space in order to avoid being confounded by Boltzmann's miasma.
Either the flux projectors of section \ref{obs}, or more general
relational constructions, could usefully play a role in this.   We
also expect considerations analogous  to those above to lead to a
finite number of states for many such spacetimes.  Investigation of these issues is in progress.

\appendix

\section{Counting Physical States}
\label{count}

In the main text, we identified a space ${\cal H}^f_{phys}$ for
which perturbative gravity is weakly coupled.   We also noted that these physical
states can be obtained  by acting with the group averaging map
$\eta$ on the states ${\cal H}^{{\cal F}, f}$ defined with respect
to a fixed foliation.  This fits with the intuition that every
physical state should have a `gauge-fixed' representative in the
original Fock space.  However, as a test of the formalism, we might
ask to what extent states in a given ${\cal H}^{{\cal F}, f}$ are
`gauge-equivalent' to each other.  That is, we may ask how much
smaller is the physical space ${\cal H}^{f}_{phys}$ as compared to
the space of `seed' states ${\cal H}^{{\cal F}, f}$. Physically,
global concerns arising from consistent coupling to gravity should
not strongly affect the local physics of low-energy states, so the
entropies associated with these two spaces should be comparable.

We now argue that these entropies are indeed comparable.  We will construct a space of seed states ${\cal H}_{seed}$ on which $\eta$ has trivial kernel, but whose entropy agrees with that of ${\cal H}^{{\cal F},f}$ to leading order in ${\Rds}/{\ell_p}$.  To proceed, recall that $\H^{f}_{phys}$ can
be obtained by group averaging any $\H^{{\cal F},f}$ and let $e^{S(f)}$ be the dimension of $\H^{{\cal F},f}$.  Consider now the entropy $S(f,J)$ of the ensemble of states in $\H^{{\cal F},f}$ having total angular momentum $J$ on the $S^{D-1}$ spheres of our foliation.  This entropy is maximized at $J=0$.  For for large $f$, we are in the thermodynamic limit and the entropy of $\H^{{\cal F},f}$ is dominated by states with $J=0$.  Thus,  $S(f, J=0) \approx S(f)$ up to a logarithmic correction.

Furthermore, note that a generic $J=0$ state will also be in {\it local} thermodynamic
equilibrium.  In particular, consider any small region of the neck
with size $(\Rds_c)$ small compared to $\Rds$.  To a good
approximation, our state in this region will approximate a thermal state in flat spacetime.  In this approximation we may say that the expected (spatial) momentum $P_i$ vanishes in this region.  Restricting our ensemble to states with $P_i=0$ in this region again makes only a logarithmic correction to the entropy.  Doing so for a set of order  $(\Rds/\Rds_c)^{D-1}$ regions which cover the neck corrects the entropy by a factor which remains subleading for $\Rds_c \gg (\Rds^2 \ell_p^{D-2})^{\frac{1}{D}}$, where we have neglected a final logarithmic correction to the condition on $\Rds_c$.   We will refer to the resulting class of states as the set ${\cal H}_{seed}$ of `locally zero-momentum seed states.'

We now argue that all locally zero-momentum seed states lead
to independent physical states under group averaging.   Consider two such seed states $\aux
\psi_1 \rangle$ and $\aux \psi_2 \rangle$ which are orthogonal in
${\cal H}_0$.  Since we required all seed states to have $J=0$, spherical symmetry may be used to reduce the group averaging inner product (\ref{gaip})  to the one-dimensional integral (see e.g. \cite{AHII})
\begin{equation}
\int d \lambda \mu(\lambda) \langle \psi_1 \aux \exp(i \lambda B) \aux \psi_2 \rangle
\end{equation}
where $\mu(\lambda) >0 $ is an appropriate measure factor and $B$ generates a `boost' which fixes some $S^{D-2}$ in the neck.   Now consider a local region in which we required our seed states to satisfy $P_i=0$, and which intersects this $S^{D-2}$.  Since the energy
density is non-zero, the boosted state $U(B) \aux
\psi_2 \rangle$ represents an eigenstate of this local $P_i$ with non-zero eigenvalue.   It is therefore orthogonal to any $P_i=0$ eigenstate such as $\aux \psi_1 \rangle$.  In particular, we have $\langle \Psi_1 \pip \Psi_2 \rangle =0$.   On the other hand, since $\mu(\lambda) > 0$ and we have argued that $\langle \psi_1 \aux \exp(i \lambda B) \aux \psi_2 \rangle $ is essentially a $\delta$-function supported at $\lambda=0$, this argument also demonstrates that the group averaging inner product (\ref{gaip}) is positive definite on this class of states.  It follows that  $\pip \Psi_1 \rangle$, $\pip  \Psi_2 \rangle$ are linearly independent in $\H^{f}_{phys}$.

Let us summarize: For sufficiently large $f$ we have found a class
of seed states which are mapped into independent physical states by
group averaging.  We have also argued that the entropy of this class
of seed states is $S(f)$ up to logarithmic corrections.  But $S(f)$ is the entropy of the full space of states ${\cal H}^{{\cal F}, f}$ that we wish to group average. It follows that $S(f)$ is also the entropy of $\H^{f}_{phys}$, so that our physical state space has the expected entropy.

\section*{Acknowledgements}

We would like to thank D. Page and B. Losic, and especially  J.
Hartle and A. Higuchi  for valuable discussions.  In particular,
appendix \ref{count} is an outgrowth of discussions with A. Higuchi.
S.G. was supported in part by the Department of Energy under
Contract DE-FG02-91ER40618, and by grant RFPI-06-18 from the
Foundational Questions Institute (fqxi.org). D.M. was supported in
part by the National Science Foundation under Grant No PHY03-54978,
and by funds from the University of California.


\begin{thebibliography}{11}

\bibitem{Banks:2000fe}
  T.~Banks,
  ``Cosmological breaking of supersymmetry or little Lambda goes back to  the
  future. II,''
  arXiv:hep-th/0007146.

\bibitem{Fischunpub}
W.~Fischler, unpublished (2000).

\bibitem{Dyson:2002nt}
  L.~Dyson, J.~Lindesay and L.~Susskind,
  ``Is there really a de Sitter/CFT duality,''
  JHEP {\bf 0208}, 045 (2002)
  [arXiv:hep-th/0202163].

\bibitem{VMLS} V. Moncrief, ``Spacetime symmetries and linearization stability of the Einstein equations. I ,'' J. Math Phys. {\bf 16} (1975) 493; Space-time symmetries and linearization stability of the Einstein equations. II,'' J. Math Phys. {\bf 17} (1976) 1893.

\bibitem{Arms} J. Arms, ``Linearization stability of the Einstein-Maxwell system,'' J. Math. Phys 18, 830 (1977); ``Linearization stability of gravitational and gauge fields,'' J. Math. Phys. 20, 443 (1979)

\bibitem{VM} V. Moncrief, ``Invariant states and quantized gravitational perturbations,'' Phys. Rev. D18 (1978) 983;  ``Quantum Linearization Instabilities,'' Gen. Rel. Grav. {\bf 10} (1978) 93.

\bibitem{FM} A. E. Fisher and J. E. Marsden, in {\it General Relativity: An Einstein Centenary Survey}, edited by S. W. Hawking and W. Israel (Cambridge University Press, Cambridge, England, 1979).

\bibitem{AMM}  J. Arms, J. Marsden and V. Moncrief, The Structure of the Space of Solutions of EinsteinÕs Equations II, Annals of Physics 144 (1982) 81.

\bibitem{AH1}
 A.~Higuchi,
  ``Quantum linearization instabilities of de Sitter space-time. 1,''
  Class.\ Quant.\ Grav.\  {\bf 8}, 1961 (1991).

\bibitem{LU} B.~Losic and W.~G.~Unruh,
  ``On leading order gravitational backreactions in de Sitter spacetime,''
  Phys.\ Rev.\  D {\bf 74}, 023511 (2006)
  [arXiv:gr-qc/0604122].

\bibitem{BanksStable} T.~Banks,
  ``More thoughts on the quantum theory of stable de Sitter space,''
  arXiv:hep-th/0503066.

\bibitem{AHII}
 A.~Higuchi,
  ``Quantum linearization instabilities of de Sitter space-time. 2,''
  Class.\ Quant.\ Grav.\  {\bf 8}, 1983 (1991).


  \bibitem{ALMMT} A.~Ashtekar, J.~Lewandowski, D.~Marolf, J.~Mourao and T.~Thiemann,
  ``Quantization of diffeomorphism invariant theories of connections with local
  degrees of freedom,''
  J.\ Math.\ Phys.\  {\bf 36}, 6456 (1995)
  [arXiv:gr-qc/9504018].


\bibitem{AHprivate} Unpublished notes from A. Higuchi.


  \bibitem{SO21}
 A.~Gomberoff and D.~Marolf,
  ``On group averaging for SO(n,1),''
  Int.\ J.\ Mod.\ Phys.\  D {\bf 8}, 519 (1999)
  [arXiv:gr-qc/9902069].

  \bibitem{JL}
  J.~Louko,
  ``Group averaging, positive definiteness and superselection sectors,''
  J.\ Phys.\ Conf.\ Ser.\  {\bf 33}, 142 (2006)
  [arXiv:gr-qc/0512076];
  ``Group averaging, positive definiteness and superselection sectors,''
  J.\ Phys.\ Conf.\ Ser.\  {\bf 33}, 142 (2006)
  [arXiv:gr-qc/0512076];
 J.~Louko and A.~Molgado,
  ``Superselection sectors in the Ashtekar-Horowitz-Boulware model,''
  Class.\ Quant.\ Grav.\  {\bf 22}, 4007 (2005)
  [arXiv:gr-qc/0505097].


\bibitem{single}
   D.~Marolf,
  ``Refined algebraic quantization: Systems with a single constraint,''
  arXiv:gr-qc/9508015.

\bibitem{KL} N.~P.~Landsman,
  ``Rieffel induction as generalized quantum Marsden-Weinstein reduction,''
  arXiv:hep-th/9305088.

\bibitem{QORD}
 D.~Marolf, Ph.D Thesis (1992);
  ``Quantum observables and recollapsing dynamics,''
  Class.\ Quant.\ Grav.\  {\bf 12}, 1199 (1995)
  [arXiv:gr-qc/9404053].



\bibitem{WhereAreWe}
  D.~Marolf,
  ``Group averaging and refined algebraic quantization: Where are we now?,''
  arXiv:gr-qc/0011112.

\bibitem{MG}
  D.~Giulini and D.~Marolf,
  ``A uniqueness theorem for constraint quantization,''
  Class.\ Quant.\ Grav.\  {\bf 16}, 2489 (1999)
  [arXiv:gr-qc/9902045].

\bibitem{OS}
    O.~Y.~Shvedov,
  ``Refined algebraic quantization of constrained systems with structure
  functions,''
  arXiv:hep-th/0107064;
  ``On correspondence of BRST-BFV, Dirac and refined algebraic  quantizations
  of constrained systems,''
  Annals Phys.\  {\bf 302}, 2 (2002)
  [arXiv:hep-th/0111270].



\bibitem{GMH}
 S.~B.~Giddings, D.~Marolf and J.~B.~Hartle,
  ``Observables in effective gravity,''
  Phys.\ Rev.\  D {\bf 74}, 064018 (2006)
  [arXiv:hep-th/0512200].



\bibitem{DMPI}   D.~Marolf,
  ``Path Integrals and Instantons in Quantum Gravity,''
  Phys.\ Rev.\  D {\bf 53}, 6979 (1996)
  [arXiv:gr-qc/9602019].

\bibitem{Hotta:1992qy}
  M.~Hotta and M.~Tanaka,
  ``Shock wave geometry with nonvanishing cosmological constant,''
  Class.\ Quant.\ Grav.\  {\bf 10}, 307 (1993).

\bibitem{Esposito:2006vv}
  G.~Esposito, R.~Pettorino and P.~Scudellaro,
  ``On the ultrarelativistic limit of boosted space-times with cosmological
  constant,''
  arXiv:gr-qc/0606126.

\bibitem{stiff}
 T.~Banks, W.~Fischler, A.~Kashani-Poor, R.~McNees and S.~Paban,
  ``Entropy of the stiffest stars,''
  Class.\ Quant.\ Grav.\  {\bf 19}, 4717 (2002)
  [arXiv:hep-th/0206096].


\bibitem{EinsDov}
A. Einstein, {\sl Relativity: the special and general theory}
(Dover, New York).

\bibitem{DeWitt:1962cg}
  B.~S.~DeWitt,
  ``The Quantization of geometry,''
in {\sl Gravitation: An introduction to current research} L. Witten
(ed.) Wiley, 1962.

\bibitem{DeWitt:1967yk}
  B.~S.~DeWitt,
  ``Quantum Theory of Gravity. 1. The Canonical Theory,''
  Phys.\ Rev.\  {\bf 160}, 1113 (1967).

\bibitem{Page:1983uc}
  D.~N.~Page and W.~K.~Wootters,
  ``Evolution Without Evolution: Dynamics Described By Stationary
  Observables,''
  Phys.\ Rev.\  D {\bf 27}, 2885 (1983).

\bibitem{Banks:1984cw}
  T.~Banks,
  ``T C P, Quantum Gravity, The Cosmological Constant And All That..,''
  Nucl.\ Phys.\  B {\bf 249}, 332 (1985).

\bibitem{Hartle:1986gn}
  J.~B.~Hartle,
  ``Prediction in quantum cosmology,''
in {\it Cargese 1986, proceedings, gravitation in astrophysics,
329-360. }

\bibitem{Rovelli:1990jm}
  C.~Rovelli,
  ``Quantum mechanics without time: a model,''
  Phys.\ Rev.\  D {\bf 42}, 2638 (1990).

\bibitem{Rovelli:1990ph}
  C.~Rovelli,
  ``What is observable in classical and quantum gravity?,''
  Class.\ Quant.\ Grav.\  {\bf 8}, 297 (1991).

\bibitem{Rovelli:1990pi}
  C.~Rovelli,
  ``Quantum reference systems,''
  Class.\ Quant.\ Grav.\  {\bf 8}, 317 (1991).

\bibitem{Rovelli:2001bz}
  C.~Rovelli,
  ``Partial observables,''
  Phys.\ Rev.\  D {\bf 65}, 124013 (2002)
  [arXiv:gr-qc/0110035].

\bibitem{Tsamis:1989yu}
  N.~C.~Tsamis and R.~P.~Woodard,
  ``Physical Green's functions in quantum gravity,''
  Annals Phys.\  {\bf 215}, 96 (1992).

\bibitem{MarT}
D.~Marolf,
  ``Almost ideal clocks in quantum cosmology: A Brief derivation of time,''
  Class.\ Quant.\ Grav.\  {\bf 12}, 2469 (1995)
  [arXiv:gr-qc/9412016].



\bibitem{Ambjorn:1996mk}
  J.~Ambjorn, K.~N.~Anagnostopoulos, U.~Magnea and G.~Thorleifsson,
  ``Geometrical interpretation of the KPZ exponents,''
  Phys.\ Lett.\  B {\bf 388}, 713 (1996)
  [arXiv:hep-lat/9606012].

\bibitem{Ambjorn:1996wc}
  J.~Ambjorn and K.~N.~Anagnostopoulos,
  ``Quantum geometry of 2D gravity coupled to unitary matter,''
  Nucl.\ Phys.\  B {\bf 497}, 445 (1997)
  [arXiv:hep-lat/9701006].






\bibitem{Gambini:2004pe}
  R.~Gambini, R.~Porto and J.~Pullin,
  ``A relational solution to the problem of time in quantum mechanics and
  quantum gravity induces a fundamental mechanism for quantum  decoherence,''
  New J.\ Phys.\  {\bf 6}, 45 (2004)
  [arXiv:gr-qc/0402118].

\bibitem{Gambini:2006yj}
  R.~Gambini, R.~Porto and J.~Pullin,
  ``Fundamental decoherence from quantum gravity: A pedagogical review,''
  arXiv:gr-qc/0603090.

\bibitem{Dittrich:2005kc}
  B.~Dittrich,
  ``Partial and Complete Observables for Canonical General Relativity,''
  Class.\ Quant.\ Grav.\  {\bf 23}, 6155 (2006)
  [arXiv:gr-qc/0507106].

  \bibitem{Thiemann:2004wk}
  T.~Thiemann,
  ``Reduced phase space quantization and Dirac observables,''
  Class.\ Quant.\ Grav.\  {\bf 23}, 1163 (2006)
  [arXiv:gr-qc/0411031].

\bibitem{Pons:2005rz}
  J.~M.~Pons and D.~C.~Salisbury,
  ``The issue of time in generally covariant theories and the Komar-Bergmann
  approach to observables in general relativity,''
  Phys.\ Rev.\  D {\bf 71}, 124012 (2005)
  [arXiv:gr-qc/0503013].




\bibitem{Gary:2006mw}
  M.~Gary and S.~B.~Giddings,
 ``Relational observables in 2d quantum gravity,''
  arXiv:hep-th/0612191, to appear in {\sl Phys. Rev. D}.

\bibitem{Boltzmann}
L. Boltzmann, ``On certain questions of the theory of gases," Nature
{\bf 51}, 413 (1895).

\bibitem{Rees:1998jp}
  M.~Rees,
  {\sl Before the beginning: our universe and others} (Simon and Schuster, New York, 1997), p 221.

\bibitem{Albrecht:2004ke}
  A.~Albrecht and L.~Sorbo,
  ``Can the universe afford inflation?,''
  Phys.\ Rev.\  D {\bf 70}, 063528 (2004)
  [arXiv:hep-th/0405270].

\bibitem{Page:2006dt}
  D.~N.~Page,
  ``Is our universe likely to decay within 20 billion years?,''
  arXiv:hep-th/0610079.

\bibitem{BF}
R.~Bousso and B.~Freivogel,
  ``A paradox in the global description of the multiverse,''
  arXiv:hep-th/0610132.

\bibitem{Hartle:2007zv}
  J.~B.~Hartle and M.~Srednicki,
  ``Are We Typical?,''
  arXiv:0704.2630 [hep-th].

\bibitem{Giddings:2006sj}
  S.~B.~Giddings,
  ``Black hole information, unitarity, and nonlocality,''
  Phys.\ Rev.\  D {\bf 74}, 106005 (2006)
  [arXiv:hep-th/0605196].

\bibitem{Giddings:2006be}
  S.~B.~Giddings,
``(Non)perturbative gravity, nonlocality, and nice slices,''
  Phys.\ Rev.\  D {\bf 74}, 106009 (2006)
  [arXiv:hep-th/0606146].

  \bibitem{Giddings:2006vu}
  S.~B.~Giddings,
  ``Locality in quantum gravity and string theory,''
  Phys.\ Rev.\  D {\bf 74}, 106006 (2006)
  [arXiv:hep-th/0604072].

\bibitem{Giddings:2007ie}
  S.~B.~Giddings,
  ``Quantization in black hole backgrounds,''
  arXiv:hep-th/0703116.


\bibitem{Giddings:2001pt}
  S.~B.~Giddings and M.~Lippert,
  ``Precursors, black holes, and a locality bound,''
  Phys.\ Rev.\  D {\bf 65}, 024006 (2002)
  [arXiv:hep-th/0103231].

  \bibitem{Giddings:2004ud}
  S.~B.~Giddings and M.~Lippert,
 ``The information paradox and the locality bound,''
  Phys.\ Rev.\  D {\bf 69}, 124019 (2004)
  [arXiv:hep-th/0402073].




\bibitem{dSAdv}
 R.~Bousso,
  ``Adventures in de Sitter space,''
  arXiv:hep-th/0205177.

\bibitem{Page:1993wv}
  D.~N.~Page,
  ``Information in black hole radiation,''
  Phys.\ Rev.\ Lett.\  {\bf 71}, 3743 (1993)
  [arXiv:hep-th/9306083].

\bibitem{NAH}
N. Arkani-Hamed, talk at the KITP conference {\sl String phenomenology 2006}.

\bibitem{Arkani-Hamed:2007ky}
  N.~Arkani-Hamed, S.~Dubovsky, A.~Nicolis, E.~Trincherini and G.~Villadoro,
  ``A Measure of de Sitter Entropy and Eternal Inflation,''
  arXiv:0704.1814 [hep-th].


\bibitem{DO}
  U.~H.~Danielsson and M.~E.~Olsson,
  ``On thermalization in de Sitter space,''
  JHEP {\bf 0403}, 036 (2004)
  [arXiv:hep-th/0309163].

\bibitem{Banks:2002wr}
  T.~Banks, W.~Fischler and S.~Paban,
  ``Recurrent nightmares?: Measurement theory in de Sitter space,''
  JHEP {\bf 0212}, 062 (2002)
  [arXiv:hep-th/0210160].

\bibitem{dStrouble}
N.~Goheer, M.~Kleban and L.~Susskind,
  ``The trouble with de Sitter space,''
  JHEP {\bf 0307}, 056 (2003)
  [arXiv:hep-th/0212209].


\bibitem{Fr}  A.~Casher, F.~Englert, N.~Itzhaki, S.~Massar and R.~Parentani,
  ``Black hole horizon fluctuations,''
  Nucl.\ Phys.\  B {\bf 484}, 419 (1997)
  [arXiv:hep-th/9606106].


\bibitem{RS}  R.~D.~Sorkin,
  ``How wrinkled is the surface of a black hole?,''
  arXiv:gr-qc/9701056.

\bibitem{width} D.~Marolf,
  ``On the quantum width of a black hole horizon,''
  [arXiv:hep-th/0312059].



\bibitem{Dyson:2002pf}
  L.~Dyson, M.~Kleban and L.~Susskind,
  ``Disturbing implications of a cosmological constant,''
  JHEP {\bf 0210}, 011 (2002)
  [arXiv:hep-th/0208013].



\end{thebibliography}
\end{document}